\documentclass[aip,rsi,amsmath,amssymb,reprint,graphicx]{revtex4-2} 
\usepackage{graphicx}
\usepackage{dcolumn}
\usepackage{bm}
\usepackage{cancel}
\usepackage{float}

\makeatother
\begin{document}
\title{High-resolution bandpass x-ray imaging with crystal reflectors: overcoming geometric aberrations}
\thanks{Submitted to Rev. Sci. Instrum}
\author{S. Stoupin}
\email[The author to whom correspondence may be addressed: ]{stoupin1@llnl.gov}
\affiliation{Lawrence Livermore National Laboratory, Livermore, CA 94550, USA} 
\author{D. Sagan}
\affiliation{Cornell University, Ithaca, NY 14853, USA}


\begin{abstract}

The imaging problem of a specular reflector is revisited. 
Retaining terms through second order in the reflector surface expansion, we derive the form of the  aberration-limiting aperture for arbitrary magnification assuming no bandwidth limitations.
A permissible relative aperture size of the reflector is limited by a set relative aberration tolerance and scales with the tangent of the central glancing angle of incidence.
These limiting aberrations become practically insignificant near backscattering.
The results extend to x-ray diffracting crystals in symmetric Bragg geometry shaped as an ellipsoid of revolution.
This geometry permits polychromatic imaging for hard x-rays over a bandwidth defined by the accepted range of Bragg angles, thereby suppressing aberrations of higher orders. 
We assess ellipsoidal crystal imagers using ray tracing simulations for two high-magnification designs with Bragg angles far from and close to backscattering.  
In both cases the ellipsoidal crystals produce images of higher quality compared to those formed by equivalent toroidal crystal imagers.
\end{abstract}

\maketitle

\section{Introduction}

Reflection-type optics is employed in x-ray instrumentation as light condensers, spectral analyzers and imagers. Imaging using reflecting optics is of particular importance 
for plasma diagnostics\cite{Forster91,Uschmann95,Pikuz97,Brown97,Koch98,Fujita01,Aglitskiy08,Stoeckl12,Schollmeier15,Hall19,Ebert25}
where detection of reflected radiation off the primary (direct) line of sight is preferable to achieve high signal/background ratios. 
Another major advantage is that x-ray reflectors typically operate within a limited band of photon energies, which enables direct energy-specific image analysis. The bandwidth is either limited by a cut-off x-ray energy (x-ray mirrors at grazing incidence) or truncated by the accessible range of Bragg angles in diffraction from crystals and multilayers. 

\begin{figure}
\includegraphics[scale=0.4]{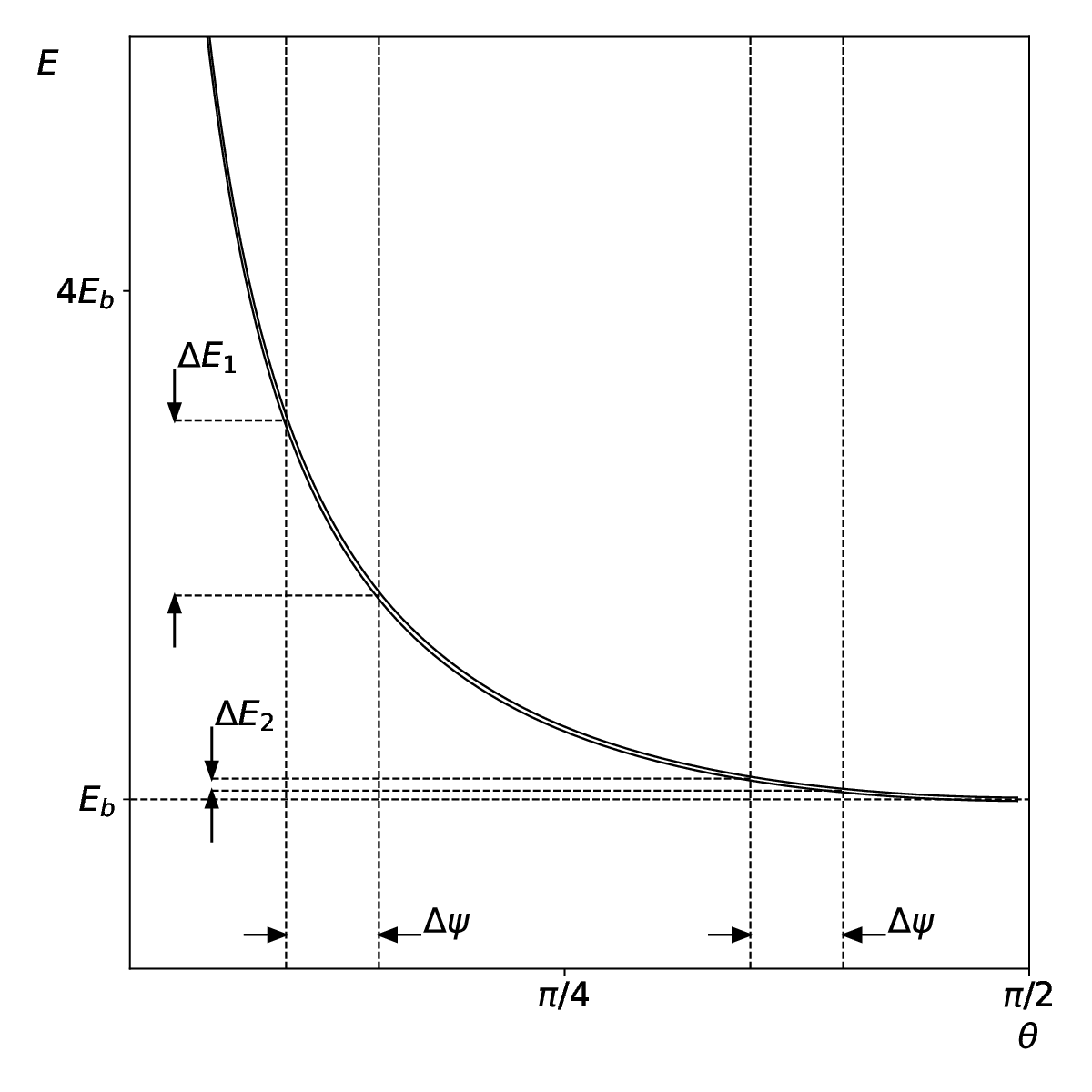}
\caption{\label{fig:dumond} DuMond diagram of a crystal reflection for the energy-angle ($E$-$\theta$) entrance phase space. 
For practical values of accepted angular divergence $\Delta \psi$ the energy bandwidth selected by the crystal reflection is dominated by the Bragg's law dispersion $\Delta E = E \Delta \psi /\tan{\theta}$ for the case of intermediate Bragg angles ($\Delta E_1$) as well as for near-backscattering case ($\Delta E_2$). The intrinsic width of the reflection region is exaggerated for clarity.}
\end{figure}

The case of crystals deserves special attention as it permits imaging at angles far from grazing incidence, which helps reducing variation in magnification for rays reflected at different angles and enables larger collection angles for optics of compact dimensions ($\lesssim 100$~mm). However, the enhancement is counteracted by the narrowness of intrinsic width of the reflection region for crystals. 
Figure~\ref{fig:dumond} shows a DuMond diagram\cite{DuMond37} for the energy-angle ($E$-$\theta$) entrance phase space of a crystal reflection. 
The solid lines are the boundaries of the reflection region $E_{\pm} = E \pm \frac{1}{2}\varepsilon_i E$, 
where $E = E_b/\sin{\theta}$ is the Bragg's law ($E_b$ being the backscattering energy of a given crystal reflection), and $\varepsilon_i$ is the relative intrinsic energy width. 
Perfect crystals of high lattice quality (e.g., Si) are of primary importance in imaging applications. For such crystals $\varepsilon_i \simeq 10^{-4} - 10^{-6}$, where the smaller values correspond to higher-order reflections. The effective intrinsic width may increase in curved crystals due to lattice deformations by about one order of magnitude.
In Fig.~\ref{fig:dumond} the intrinsic width is further increased for clarity ($\varepsilon_i = 10^{-2}$). 
For practical values of accepted angular divergence $\Delta \psi \simeq 0.01 - 0.1$ the energy bandwidth selected by a crystal reflection is dominated by the Bragg's law dispersion $\Delta E = E \Delta \psi/\tan{\theta}$. This is illustrated in Fig.~\ref{fig:dumond} for the case of intermediate Bragg angles ($\Delta E_1$) and for the near-backscattering case ($\Delta E_2$). 

Commonly used x-ray crystal imagers have either spherical or toroidal shapes, which exhibit reflector aberrations of different orders in the surface expansion originating from variations in the optical path connecting arbitrary chosen set of object, reflector, and image points\cite{Noda74,Howells94,Koch98,Schollmeier15}. Toroidally-shaped x-ray crystals are used to correct second-order astigmatic aberration in imaging of a point source by matching meridional and sagittal paraxial focal distances. \cite{Forster91,Uschmann95,Koch98,Schollmeier15,Ebert25} 
In general, the second-order aberrations as well as those of higher orders remain in imaging an extended object. 
A practical workaround in reducing the aberrations is to limit the size of the crystal to an extent where the accepted bandwidths (due to the accessible range of Bragg angles) are reduced to a few eV.
Meanwhile, relative bandwidths for x-ray emission of interest are frequently $\Delta E_X \simeq 10 - 100$~eV (e.g. plasma emission), which may lead to photometric limitations in applications of 
crystal imagers. 

Another well-known shape for an aspherical reflector is ellipsoid of revolution for which a point source placed in one of its foci forms an aberration-free image at the other focus. \cite{KB48,Berreman77,Authier}
An x-ray reflecting crystal with working atomic planes parallel to its surface (i.e. symmetric reflection) conforming to the ellipsoidal surface approximates such an imager as long as penetration of reflected x-rays into the bulk of the crystal is considered to be small. Such curved crystal can be understood as a special case of variable-radii sagittally-focusing elliptical spectrometer \cite{Stoupin24} where the local rotation axes forming the surface are aligned with the major axis of the ellipse.  
The basic property of the ellipsoid of revolution is that the focus-to-focus path length is independent on the angle of incidence to the inner surface. 
An image produced by such crystal reflector is thus polychromatic as it is formed by rays reflected at different Bragg angles. 
This can lead to a larger energy bandwidth accessible for imaging which enables enhancement in the photon flux delivered to a detector. 
However, perfect focus-to-focus imaging for an ellipsoidal geometry (and, in general, any fixed imaging geometry) is achieved only for a source at its nominal position. Displacing the source introduces defocus and aberrations. 
Image aberrations become severe for regions of interests at the periphery thus limiting useful field of view. As argued by Kirkpatrick and Baez \cite{KB48} the useful field of view by two crossed mirrors far exceeds that of a perfectly ellipsoidal mirror. Other sequential arrangements of mirrors such as Wolter type\cite{Wolter52} are known to reduce aberrations. It is, however, not straightforward to adopt these concepts in x-ray crystal optics because crystals select a rather narrow reflection region from the incident divergent and polychromatic radiation. A multi-crystal imaging arrangement requires a dispersion match for rays forming the image to be subsequently reflected. Furthermore, non-sequential (one-element) imagers are preferable as diagnostic instruments because of relative ease of their implementation and alignment.\cite{Koch98} 

This study first explores how second-order aberrations treated collectively for an object of a finite size (as opposed to a point source) limit the useful field of view of a generalized specular reflector while assuming no bandwidth limitations (Sec.~\ref{sec:rayab}). Mielenz \cite{Mielenz74} derived the aberration-limited form of the reflector for an ellipsoid of revolution of unit magnification. 
A similar form is obtained in this work for a reflector of arbitrary magnification assuming that the reflector's surface satisfies point-to-point imaging condition to the second order in small deviations from its center (Sec.~\ref{sec:lform}). The permissible acceptance aperture size of the reflector scales with the paraxial focal distance, depends on the working angle of incidence and is limited by a chosen relative aberration tolerance. For cases of high magnification relevant to imaging small objects, this condition restricts the collection solid angle, which in turn limits photometric efficiency of the reflector.
In near-normal incidence the field of view is not restricted by the second-order aberrations. 
Section~\ref{sec:pm} discusses the photometrics of x-ray crystal imagers, the only class of reflectors that offer an unrestricted angle of incidence for hard x-rays.

The second part of this study presents a series of ray tracing simulations of selected crystal geometries to explore limitations and 
potential advantages in x-ray imaging with a finite band of photon energies selected by the range of accessible Bragg angles (Sec.~\ref{sec:rt}). 
We demonstrate that using aberration-limiting apertures such polychromatic bandpass imaging can be accomplished at intermediate Bragg angles far from near-normal incidence. 
In approaching near-normal incidence narrower bands of photon energies are selected by a Bragg reflection. 
Since the field of view is not restricted by the second order aberrations this regime highlights the effect of higher orders, 
which are explored by the simulations. 
In both cases, reflectors with ellipsoidal shapes were found to outperform their equivalent 
(i.e., constructed using the same local radii of curvature) toroidal crystals by producing polychromatic bandpass images of higher quality.
Consequently, higher order aberrations, termed "chromatic" for crystal reflectors, can be suppressed by choosing an ellipsoidal shape. 
In the final section of the manuscript (Sec.~\ref{sec:viab}), we discuss the viability of the practical implementation of ellipsoidal crystals,
noting that, while typical tolerances in precision surface manufacturing are sufficient to achieve the desired substrate shape, 
the main challenge is ensuring that the crystal lattice conforms to this shape.

\section{Calculation of ray aberrations}\label{sec:rayab}

\begin{figure*}[!t]
\vspace{-5cm}
\includegraphics[scale=0.8]{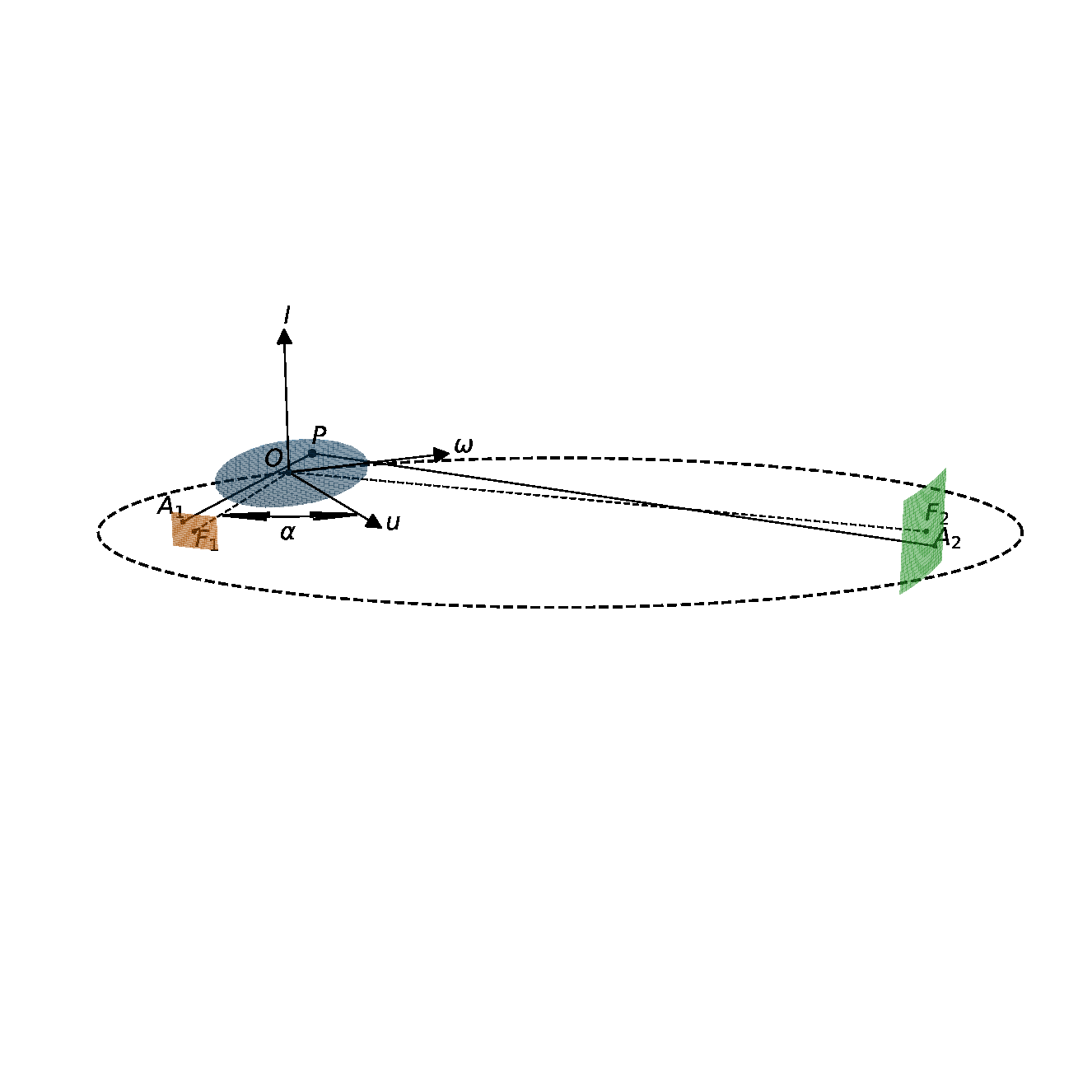}
\vspace{-7cm}
\caption{\label{fig:layout} Ellipsoidal reflector for arbitrary magnification: 
an object to be imaged and its Gaussian image are centered at $F_1$ and $F_2$, respectively. 
An arbitrary point on the reflector $P$ is defined in the local Cartesian system of coordinates [$u$,$\omega$,$l$] with origin O at the center of the reflector.
The object and the image are assumed to be small portions of cylindrical surfaces centered at O and oriented normal to the central rays $F_1O$ and $OF_2$. 
The coordinates of the object point $A_1$ and its image $A_2$ are defined in the local system using their lateral $\epsilon$ and vertical $\eta$ angular distances.
The limiting form of the reflector's aperture is obtained by a requirement that the lateral and vertical components of angular aberrations $\Delta \epsilon$ and $\Delta \eta$ combined in quadrature do not exceed a fixed value $\Delta \rho$.
}
\end{figure*}

We proceed by analogy to the derivation of ray aberrations by Mielenz \cite{Mielenz74} for an ellipsoid of revolution of unit magnification. 
An object to be imaged and its image are located at $F_1$ and $F_2$, respectively (Fig.~\ref{fig:layout}).
For magnification $M \neq 1$ the object-to-reflector central-ray distance $p$ ($F_1O$) and the reflector-to-image distance $q$ ($OF_2$) are not equal to each other ($M = q/p$). 
The reflector's surface is approximated locally to the second order in Maclaurin expansion, i.e. 
\begin{equation}\label{eq:surf}
u = \frac{\omega^2}{2R_m} + \frac{l^2}{2R_s} + ...,
\end{equation}
where $u$, $\omega$ and $l$ are the Cartesian coordinates in the local system with origin $O$ at the center of the reflector. 
For an arbitrary point $P$ on the reflector [$u$,$\omega$,$l$] are the normal, meridional and sagittal coordinates, respectively. 
The coefficients $R_m$ and $R_s$ are the meridional and sagittal local radii of curvature. 
The above Eq.~\eqref{eq:surf} is accurate to within the third order for a toroidal surface and for an ellipsoid of revolution at its vertices. 
In a general case (arbitrary placement of reflector on ellipsoid of revolution) third-order terms are present.\cite{Goldberg22}
In this regard, our consideration is not restricted to a particular overall geometry of the reflector. 
By way of example a layout where the reflector is a section of an ellipsoid of revolution is shown in Fig.~\ref{fig:layout}.

Imaging of a point source is attained using equal paraxial focal lengths $f = pq/(p+q)$ in the meridional and the sagittal planes. 
The required radii of curvature follow from Coddington's equations\cite{Schollmeier15,Goldberg22,Ebert25}: 
\begin{subequations}\label{eq:R}
\begin{align}
R_m &= \frac{2f}{\sin{\theta}} \label{eq:sub:a} \\
R_s  &= 2f\sin{\theta} \label{eq:sub:b}
\end{align}
\end{subequations}
where $\theta$ is the central-ray glancing angle of incidence to the reflector to be distinguished from the angle of incidence $\alpha = \pi/2 - \theta$ in between the central ray and the normal to the reflecting surface at the origin.  

The object and its image are assumed to be small portions of cylindrical surfaces centered at $O$ and oriented normal to the central rays. 
If $\epsilon$ and $\eta$ are the lateral and vertical angular distances of the object $A_1$ and its Gaussian image $A_2$ their $[u,\omega,l]$ coordinates are 
\begin{subequations}\label{eq:coord}
\begin{align}
A_1 &= p [\cos{(\alpha + \epsilon), -\sin{(\alpha + \epsilon)}}, \tan{\eta}]     \\  
A_2 &= q [ \cos{(\alpha + \epsilon), \sin{(\alpha + \epsilon)}}, -\tan{\eta}]
\end{align}
\end{subequations}

The Hamiltonian point characteristic\cite{BornWolf99} for the three points $A_1$, $A_2$, and $P$ is 
\begin{subequations}\label{eq:main}
\begin{align}
V = A_1P + PA_2, \label{eq:sub:a} \\
A_1P = \Big[ (u - p\cos{(\alpha +\epsilon)})^2 + (\omega+p\sin{(\alpha +\epsilon)})^2 \\  \nonumber \label{eq:sub:b}  + (l - p\tan{\eta})^2 \Big]^{1/2}  \\ 
PA_2 = \Big[ (u - q\cos{(\alpha +\epsilon)})^2 + (\omega-q\sin{(\alpha +\epsilon)})^2 \\ \nonumber \label{eq:sub:c} + (l + q\tan{\eta})^2 \Big]^{1/2} 
\end{align}
\end{subequations}

To the second order in $\omega$ and $l$ (for details see Appendix \ref{sec:a1}):
\begin{subequations}\label{eq:V123}
\begin{equation}\label{eq:V123:a}
V \simeq (p+q)\sec{\eta} + V_1 \frac{\omega^2}{2f} + V_2\frac{\omega l}{2f} + V_3\frac{l^2}{2f},
\end{equation}
with
\begin{align}
V_1 &= \cos{\eta} \bigg[1 - \cos{(\alpha + \epsilon)}\cos{\alpha} - \frac{\sin^2{(\alpha + \epsilon)}}{\sec^2{\eta}} \bigg], \label{eq:V1:b} \\ 
V_2 &= 2\cos^3{\eta} \sin{(\alpha + \epsilon)} \tan{\eta} \label{eq:V2:c}, \\ 
V_3 &= \cos{\eta} \bigg[ 1 - \frac{\cos{(\alpha+\epsilon)}}{\cos{\alpha}} - \sin^2{\eta} \bigg]. \label{eq:V3:d}
\end{align}
\end{subequations}
In the above equations we used $\theta = \pi/2 - \alpha$. 
Expanding the trigonometric functions to the first order in $\epsilon$ and $\eta$ we obtain 
\begin{subequations}\label{eq:V123lin}
\begin{align}
V_1 \simeq - \epsilon \cos{\alpha}\sin{\alpha}, \\
V_2 \simeq 2 \eta \sin{\alpha}, \\
V_3 \simeq \epsilon \tan{\alpha},
\end{align}
\end{subequations}
which is similar to the result of Mielenz\cite{Mielenz74}, except for the sign of $V_2$ due to our convention of placing the source $A_1$ at negative $\omega$ coordinate. 
Further considerations and the results are identical to those in Mielenz's work upon substitution of the ellipse semi-major axis $b$ with $2f$. 
We note that the original work contains some typos (e.g. use of $l$ instead of 1). Thus, it is worth to repeat the rest of the derivation. 

According to Fermat's principle the physical ray trajectory satisfies $\partial V/ \partial \omega = \partial V/ \partial l = 0$.
The point characteristic $V$ defined above to the second order in $\omega, l$ (i.e. in the sense of Gaussian optics) satisfies Fermat's principle only to the first order in $\omega, l$. 
The residual departure of these partial derivatives from zero represent angular ray aberrations. 
The lateral and vertical components of the angular aberrations relative to the object (or image) patch are: 

\begin{subequations}\label{eq:aber}
\begin{align}
\Delta \epsilon &=\frac{1}{\cos{\alpha}} \frac{\partial V}{\partial \omega} = \frac{\omega V_1}{f \cos{\alpha}} + \frac{l V_2}{2f \cos{\alpha}} \\
\Delta \eta &= \frac{\partial V}{\partial l}  = \frac{\omega V_2}{2f} +\frac{l V_3}{f}
\end{align}
\end{subequations}

These set to a specified level can be related to a limiting form and size of the reflector's aperture as follows in the next section.

\section{Limiting Form of Reflector's Aperture and its Interpretation}\label{sec:lform}

The combined magnitude of the angular ray aberrations taken in quadrature $(\Delta \epsilon^2 + \Delta \eta^2)^{1/2}$ should not exceed a fixed value $\Delta \rho$. 
The maximum permitted values of $\omega$ and $l$ are defined by
\begin{align}\label{eq:ell}
\bigg(\frac{V_1^2}{\cos{\alpha}^2}  + \frac{V_2^2}{4} \bigg) \frac{\omega^2}{f^2} + V_2 \bigg( \frac{V_1}{\cos^2{\alpha}} + V_3 \bigg)\frac{\omega l}{f^2} \nonumber \\
+ \bigg( \frac{V^2_2}{4 \cos^2{\alpha}} + V^2_3 \bigg)\frac{l^2}{f^2} = \Delta \rho^2. 
\end{align}

\begin{figure}[!t]
\includegraphics[scale=0.45]{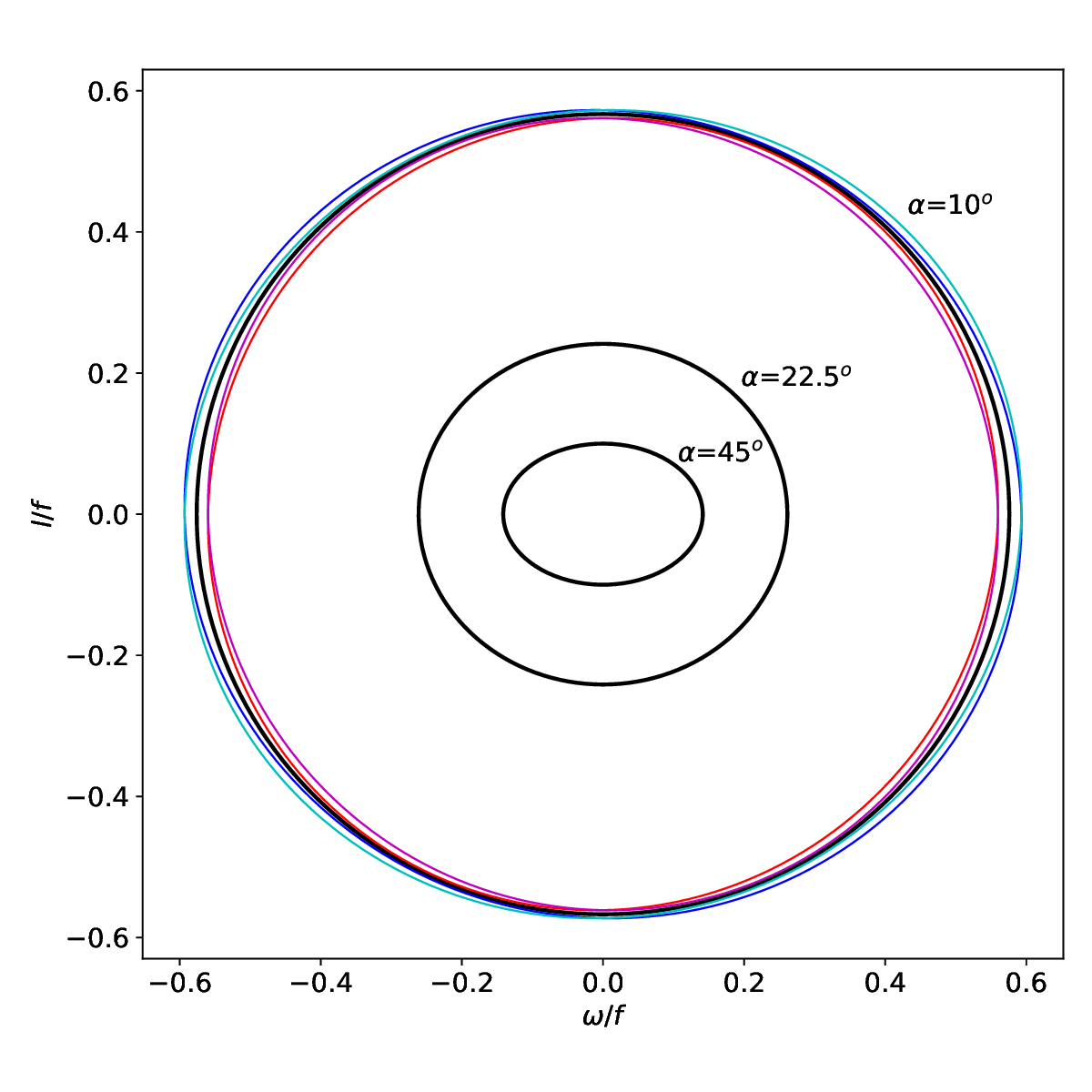}
\caption{\label{fig:ae} 
Limiting apertures of aspherical reflectors for aberration tolerance $\Delta \rho/\rho = 0.1$ and angles of incidence $\alpha~=~10^o, 22.5^o, 45^o$ by Eqs.~\ref{eq:ell0} (black). 
For the case $\alpha = 10^o$ several ellipses are plotted using more precise Eqs.~\ref{eq:V123} assuming an angular field of view with $\rho = 0.1$. These ellipses (colored) correspond to the outermost points along the diagonals across a circular field of view $\epsilon = \pm \rho/\sqrt{2}, \eta = \pm \rho/\sqrt{2}$.
}
\end{figure}

This equation defines a family of ellipses in the $\omega l$ plane with ellipse parameters depending on the angular location $\epsilon$, $\eta$ of a given object point $A_1$. 
When approximate expressions Eqs.~\eqref{eq:V123lin} for $V_1$,$V_2$ and $V_3$ are used the mixed term in Eq.~\eqref{eq:ell} vanishes and the limiting form is: 

\begin{subequations}\label{eq:ell0}
\begin{align}
(\omega/\omega_0)^2 + (l/l_0)^2 = 1, \label{eq:ell0:a} \\
\omega_0 = \frac{f}{\sin{\alpha}} \frac{\Delta \rho}{\rho}, \label{eq:ell0:b} \\
\l_0 = \frac{f}{\tan{\alpha}} \frac{\Delta \rho}{\rho}, \label{eq:ell0:c} 
\end{align}
\end{subequations}
where 
\begin{equation}\label{eq:rho}
\rho = (\epsilon^2 + \eta^2)^{1/2}.
\end{equation}
These ellipses are aligned with their major and minor axes parallel to the meridional and sagittal axes of the reflector (Fig.~\ref{fig:layout}).
For a given aberration tolerance $\Delta \rho/\rho$ Eqs.~(\ref{eq:ell0:b},\ref{eq:ell0:c}) define the limiting sizes of the reflector's semi-axes for a particular focal distance $f$. 
A few of these limiting ellipses are shown with black circles in Fig.~\ref{fig:ae} (note the normalized axis coordinates) for $\Delta \rho/\rho~=~0.1$ and $\alpha~=~10^{\circ}, ~22.5^{\circ},~45^{\circ}$, as in Ref.\cite{Mielenz74}.
The accuracy of approximate relations Eqs.~\eqref{eq:V123lin} has been checked against more precise Eqs.~\eqref{eq:V123}, where a limiting ellipse can be ascribed to each individual point of the object with angular coordinates $\epsilon, \eta$. Several such ellipses corresponding to values $\epsilon = \pm \rho/\sqrt{2}, \eta = \pm \rho/\sqrt{2}$ are shown in Fig.~\ref{fig:ae} by color for the case $\alpha~=~10^{\circ}$. A large (in practical sense) angular field of view with $\rho~=~0.1$ is assumed. In the two other cases such ellipses nearly match the main limiting curves as determined by Mielenz. It was argued that the accuracy of the approximate equations is adequate. 
We can add that for small angles of incidence (i.e., near-normal incidence) the size of the limiting ellipses becomes comparable to the focal distance $f$, exceeding the bounds of most practical applications. Consequently, the exact shape of the limiting curves becomes a secondary consideration. 

Expressing Eqs.~(\ref{eq:ell0:b},\ref{eq:ell0:c}) via the glancing angle of incidence $\theta$ one can write
\begin{subequations}\label{eq:sizes}
\begin{align}
\frac{\omega_0 \sin{\theta}}{f} = \frac{\Delta \rho}{\rho} \tan{\theta}, \\
\frac{l_0}{f} = \frac{\Delta \rho}{\rho} \tan{\theta}. 
\end{align}
\end{subequations}

One can easily recognize that the right hand sides of Eqs.~\ref{eq:sizes} represent the relative acceptance half-apertures of the reflector. 
The case of a large magnification $M~=~q/p~>>~1$, $f = pM/(1+M) \simeq p $ is of primary interest for imaging of small objects (e.g., capsule implosions at laser plasma facilities). 
The relative apertures become effectively the angular acceptances $\Delta \psi$ of the reflector in the meridional and sagittal planes. 
We thus arrive to a simple "rule-of-thumb":
\begin{equation}\label{eq:rule}
\Delta \psi = 2\frac{\Delta \rho}{\rho}  \tan{\theta}, 
\end{equation}
which states that for a set aberration tolerance $\Delta \rho/\rho$ the permissible acceptance angle of the reflector scales with tangent of the glancing angle of incidence. 
As the angle approaches near-normal incidence the aberrations are minimized to an extent that higher-order terms of surface expansion Eq.~(\ref{eq:surf}) may define the permissible maximum size of the reflector.  
An alternative, practically appealing, interpretation for Eq.~(\ref{eq:rule}) is that
for a set image resolution $\Delta \rho$ the field of view of an imaging reflector limited by aberrations scales with tangent of the glancing angle of incidence and is inversely proportional to to the reflector's relative aperture (or its angular acceptance for the case of large magnification).

\section{Relevance to photometrics of x-ray crystal imagers}\label{sec:pm}

For incident x-rays with spectral intensity $S(E)$ [photons/eV/sr] the sensitivity of a reflecting crystal is defined as $G(E) = D(E)/S(E)$, where $D(E)$ [photons/mm$^2$] is the intensity in the imaging plane.
The sensitivity [eV sr/mm$^2$] is given by (e.g., Ref.\cite{Marrs14})
\begin{equation}
G(E) = R_{\theta}\frac{dE}{d\theta}\frac{d \Omega}{dA},
\label{eq:ge}
\end{equation}
where $R_{\theta}$ is the angular-integrated reflectivity of the working crystal reflection, $dE/d\theta$ is the Bragg's law dispersion and $d\Omega/dA$ is the incremental accepted solid angle per corresponding (projected) incremental area resolution element on the detector. For these dependence on the photon energy $E$ is assumed and omitted for clarity. 
We assume that the crystal forms a polychromatic image of a uniformly emitting object with linear sizes $s_x$ and $s_y$ ($xy$ being the plane of the object normal to the central incident ray) with an accepted spectral bandwidth $\Delta E$ centered around photon energy $E_0$. The increments of solid angle $d \Omega$ are then projected on the same image area $A = M^2 s_x s_y$. 
The energy-integrated image signal in the detector plane is
\begin{equation}
D = \frac{1}{A}\int_{\Delta E}{S(E) R_e \frac{d\Omega}{dE}dE},
\label{eq:dint}
\end{equation} 
where $R_e = R_{\theta} dE/d \theta$ is the energy-integrated reflectivity. 
For a sufficiently small interval of energies the quantities in Eq.~\ref{eq:dint} can be assumed energy-independent.
The sensitivity becomes
\begin{equation}
G(E_0) \simeq R_e \frac{\Delta \Omega}{M^2 s_x s_y}, 
\label{eq:g0}
\end{equation}
where  $M$ is the magnification and $\Delta \Omega$ is the total solid angle accepted by the crystal (collection solid angle). The energy-intergrated reflectivity can be approximated as $R_e \simeq E_0 \varepsilon_i$ for crystals with low photoabsorption. 

At first glance to Eq.~\ref{eq:g0} one may arrive to a conclusion that for a given photon energy, fixed magnification and collection angle higher sensitivities can be achieved via the choice of lower order reflection, far from the near-backscattering regime, since $\varepsilon_i  \propto 1/E^2_b$ ($E_b = hc/(2d_{hkl})$, where $d_{hkl}$ is the lattice spacing associated with Miller indicies $hkl$).\cite{Shvyd'ko_book}
However, avoidance of aberrations that limit the field of view (for high $M$) requires $\Delta \Omega  \simeq (\Delta \psi)^2$ where the latter is given by Eq.~\ref{eq:rule}. Upon substitution into Eq.~\ref{eq:g0} we obtain 

\begin{equation}
G(E_0) \propto \frac{E_0}{E^2_b} \bigg(\frac{\Delta \rho}{\rho}\bigg)^2 \tan^2{\theta_0},  
\end{equation}
which, using the Bragg's law, can be reduced to 
\begin{equation}
G(E_0) \propto \frac{1}{E_0\cos^2{\theta_0}} \bigg(\frac{\Delta \rho}{\rho}\bigg)^2
\end{equation}
Thus, for a set design energy $E_0$ and aberration tolerance $\Delta \rho/\rho$ the choice of near-backscattering reflections (as $\theta$ is approaching $\pi/2$) leads to an improved photometric response.   
This general trend may have certain exceptions in practice, such as for reflections of exceptionally high reflectivity (e.g. deformation-induced enhancements of integrated reflectivity) and/or when practical geometry constraints dictate use of Bragg angles far from backscattering. 
For such cases, polychromatic x-ray imaging using aberration-limiting apertures can be used as illustrated in the following section by ray-tracing. 

An alternative way to define response of a polychromatic imager is using radiance of the imaged source object $L(E) = S(E)/(s_x s_y)$ [phot/sr/eV/mm$^2$].
It easily follows from Eq.~(\ref{eq:g0}).
\begin{equation}
D(E_0) \simeq G'(E_0) L(E_0),
\end{equation}
where 
\begin{equation}
G'(E_0) \simeq  R_e \frac{\Delta\Omega}{M^2}  
\label{eq:ge_}
\end{equation}
is the sensitivity of the imager in units [eV $\mu$sr], independent of the source dimensions. 

\section{Ray tracing simulations}\label{sec:rt}

Two x-ray imagers using Bragg reflections in Si crystal were designed to represent the two cases of intermediate and near-backscattering Bragg angles as depicted in Fig.~\ref{fig:dumond}.  
The parameters of the two imagers are summarized in Table~\ref{tab:param}.
The semi-major $a$ and semi-minor $b$ axes for the ellipsoid of revolution are related to $p$ and $q$ as follows.\cite{Goldberg22}

\begin{subequations}\label{eq:ell_ab}
\begin{align}
a & = \frac{p+q}{2} \\
b & = \sqrt{pq} \sin{\theta_0}.
\end{align}
\end{subequations}

Numerical studies were performed using Lux, a ray tracing program based upon the Bmad toolkit for charged-particle and x-ray simulations.\cite{Sagan06}
The program used crystal reflectivity tables constructed from results of calculations using pyTTE code\cite{Honkanen21},
which solved Takagi-Taupin equations for toroidally deformed Si crystals (for details see Appendix \ref{sec:a2}). 
In practical terms, the choice of Si facilitates high intrinsic crystal quality and uniformity of crystal response. 
An assumption is made that reflectivity of an ellipsoidal crystal locally is nearly identical to that of a toroidal crystal having the same principal radii of curvature. 
To explore the benefits of polychromatic imaging provided by the ellipsoidal geometry each of the two imagers was tested assuming either an ellipsoidal or toroidal crystal shapes. 
The shapes were modeled using Maclaurin surface expansions for an ellipsoid of revolution as derived by Goldberg \cite{Goldberg22} and for a torus as described in the Bmad documentation.\cite{bmad_manual} 
These expansions are included for convenience in Appendix~\ref{sec:a3}.
Parameters of the ray-tracing simulations were kept unchanged among the ellipsoidal and toroidal versions with an exception of the elliptical limiting apertures which were adjusted as described in the following. 

\begin{table*}
\caption{\label{tab:param} Design parameters of the two spectrometers studied with ray-tracing simulations: \\
$E_b = hc/(2d_{hkl})$ - backscattering energy of the reflection $hkl$, \\
$E_0$ - design energy (energy of central ray),\\
$\theta_0$ - Bragg angle for the central ray,\\
$p$ - source-to-crystal distance for the central ray (i.e., working distance),\\
$q$ - crystal-to-detector distance for the central ray,\\
$a$ - semi-major axis of the ellipsoid of revolution, \\
$b$ - semi-minor axis of the ellipsoid of revolution, \\
$M = q/p$ - nominal magnification,\\
$R_m$ - meridional radius of curvature,\\
$R_s$ - sagittal radius of curvature,\\
$\Delta E_X$ - total acceptance bandwidth,\\
$\omega_0$ - half-aperture of the crystal in the meridional direction,\\
$l_0$ -half-aperture of the crystal in the sagittal direction,\\
$G'(E_0)$ - sensitivity per Eq.~\ref{eq:ge_}.
}
\vspace{0.5cm}
\begin{tabular}{l c c c c c c c c c c c c c c}
\hline
\hline
Reflection   & $E_b$    & $E_0$     & $\theta_0$  & $p$   & $q$  & $a$     & $b$    & $M$      & $R_m$  & $R_s$      & $\Delta E_X$  & $\omega_0$   & $l_0$ &  $G'(E_0)$ \\   
                 &  [keV]    & [keV]      & [deg]          &  [m]  &  [m]  & [m]     & [m]    &             & [m]       &   [m]       & [eV]               & [mm]             & [mm] & [eV $\mu$sr] \\
\hline
Si 331        &  4.9754  & 13.000   & 22.503        & 0.5  & 5.636 &  3.068  & 0.642 & 11.272  & 2.3999   &  0.35155  & 500                & 25                  & 10      & 7.8 \\
Si 862        & 11.6405 & 11.652   & 87.455        & 0.5  & 6.498 &  3.499  & 1.800 & 12.996  & 0.92946  & 0.92763  & 20                  & 30                  & 20      & 3.9 \\
\hline
\hline
\end{tabular}
\end{table*}

\begin{figure}
\includegraphics[scale=0.35]{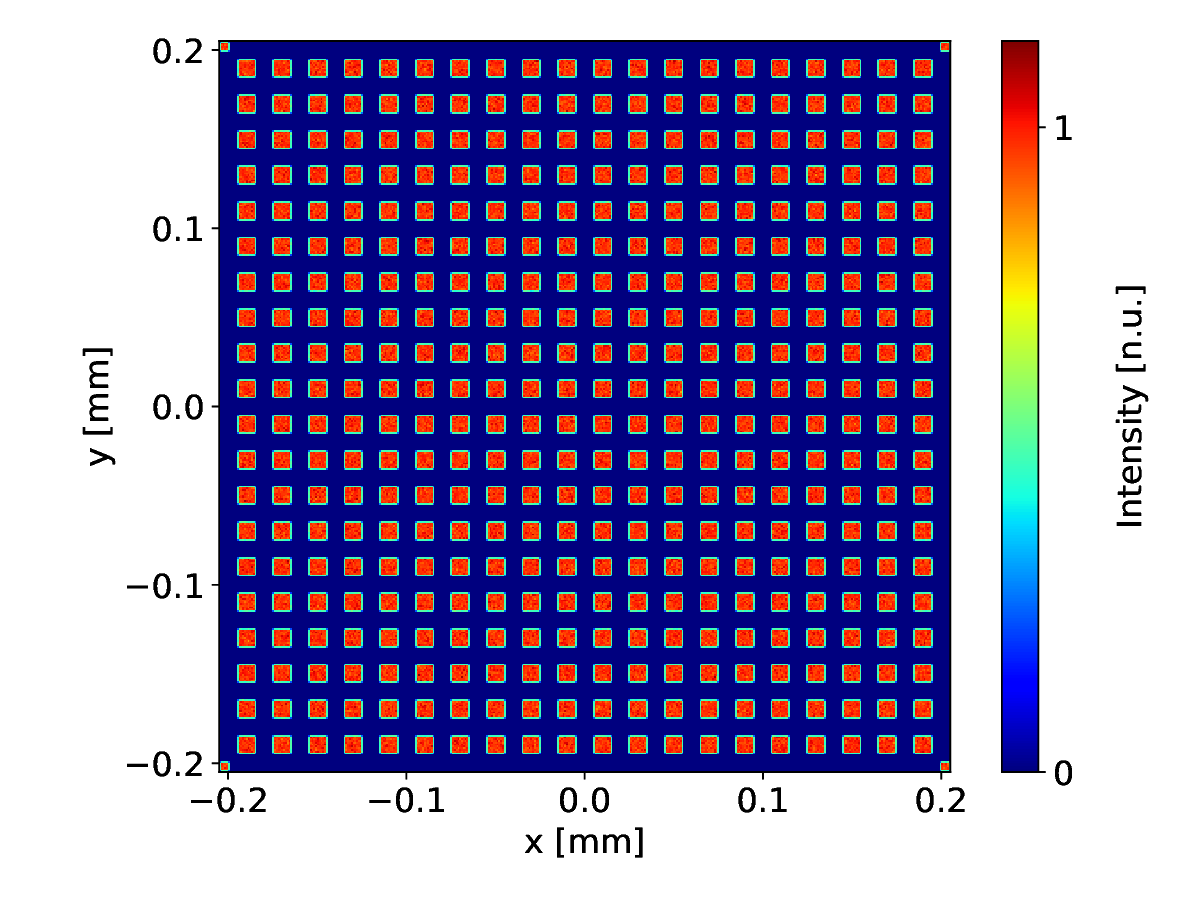}
\caption{\label{fig:mask} Mask used in the numerical simulations: a square grid with a period of 20$\mu$m; the size of 
transparent sections (squares) is 10$\mu$m. The image is obtained in a simulation where the detector is placed next to the mask with zero offset. The size of the detector pixels is 1$\times$1~$\mu$m$^2$.}
\end{figure}

The x-ray source in the simulations was a uniformly emitting square of size 0.41~$\times$~0.41 mm$^2$. The source illuminated a mask in a form of a square grid with a period of 20~$\mu$m having the size of 
fully transparent features (squares) of 10~$\mu$m. The mask was placed next to the source without an offset.  
The illuminated mask is visualized in Fig.~\ref{fig:mask}. The image is obtained using a simulation without a crystal imager and with the detector positioned right next to the mask (zero separation). The features localized to the corners are artifacts due to numerical grid construction, yet they serve as extra details to be resolved in the following simulations of crystal imagers.   
Additional ray tracing simulations were performed for each tested reflector geometry using a $ 5 \times 5 \mu m^2$ (FWHM) Gaussian source placed in the center of the object plane. The spectral and angular characteristics of the Gaussian sources were identical to those used in the main simulation with a mask illuminated by the uniform source. 
The objective of these additional simulations was to gain insights on the aberration broadening of the imaging point spread function. 

\begin{figure*}
\includegraphics[scale=0.5]{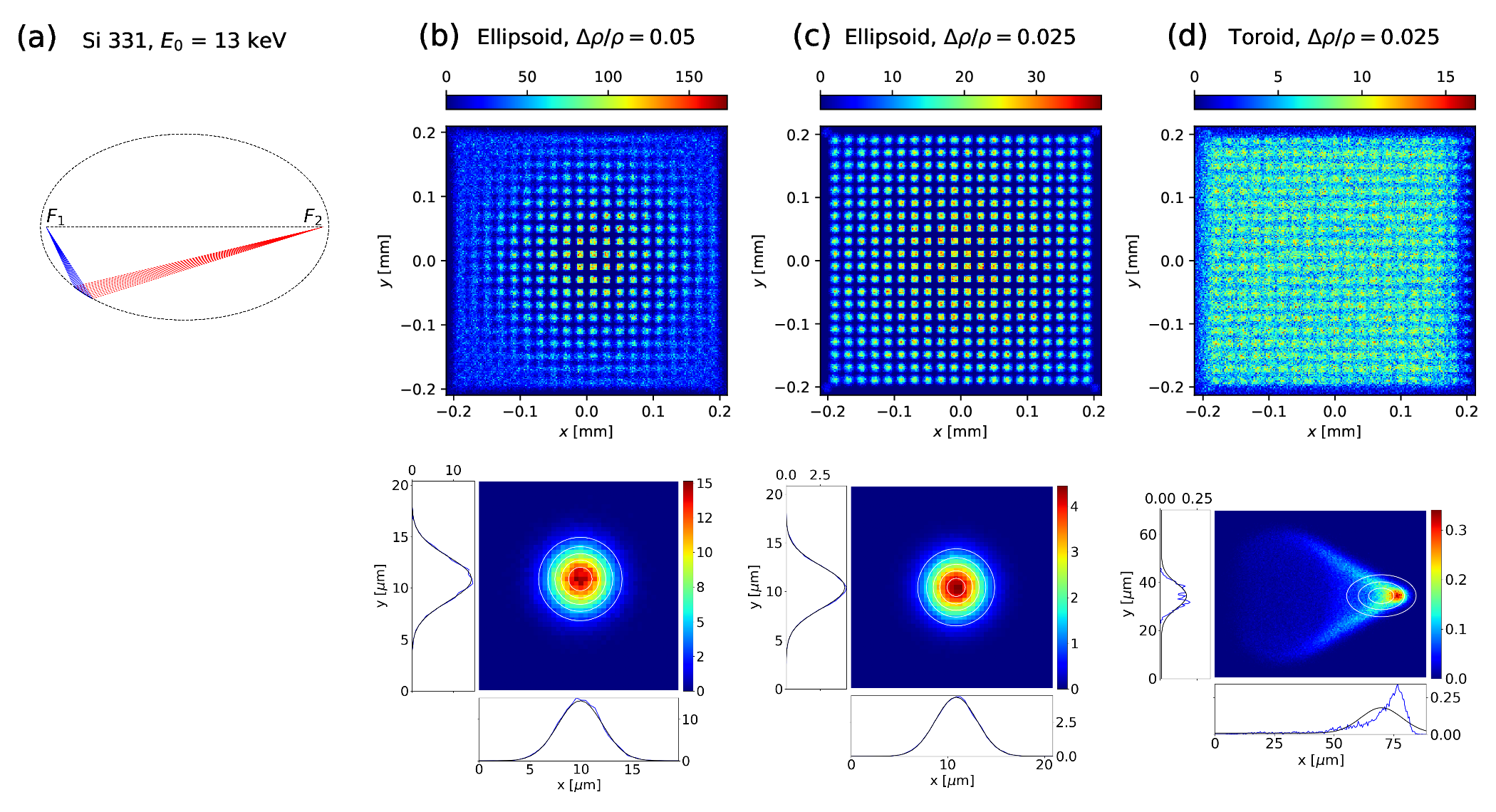}
\caption{\label{fig:Si331} 
(a) Layout of simulated ellipsoidal geometry using Si 331 reflecting crystal in the meridional midplane (not to scale). 
The x-ray source plane is located at $F_1$ normal to incident rays (blue); the detector plane is placed at $F_2$ normal to reflected rays (red).
(b)  Ray-traced images using ellipsoidal reflector with aberration-limiting apertures and photon energy range set to aberration tolerance $\Delta \rho/\rho = 0.05$ (parameters are shown in Table~\ref{tab:param}). 
(c)  Ray-traced images using ellipsoidal reflector with aberration-limiting apertures and photon energy range reduced by factor of 2 and thus set to $\Delta \rho/\rho = 0.025$.
(d) Ray-traced image using an equivalent toroidal reflector with the reduced aberration-limiting apertures  and photon energy range as in (c). 
Images in (b-c) are shown in the source coordinates $x,y$ (i.e., coordinates normalized by the nominal magnification). 
For each case the main figure (top) corresponds to a simulation using a uniform x-ray source with a mask (Fig.~\ref{fig:mask}), and the supplemental figure (bottom) corresponds to a simulation 
using a $5~\times~5~\mu m^2$ (FWHM) Gaussian source placed in the center of the object plane. The sources have uniform x-ray divergences and a uniform distribution of photon energies within the specified bandwidths. 
}
\end{figure*}

Figure~\ref{fig:Si331} summaries results of simulations for the case of intermediate working angle (Si 331). The ellipsoidal geometry is shown schematically (not to scale) in Fig.~\ref{fig:Si331}(a) in the meridional midplane. The images obtained in ray tracing simulations are shown in Figs.~\ref{fig:Si331}(b-d) in the source coordinates (i.e., the dimensions are normalized by the nominal magnification). 
For each tested geometry a supplementary image obtained in a simulation using the Gaussian source is shown below the main image simulated using the masked uniform source. 
Two-dimensional Gaussian fit is performed for the former. Each of the supplementary images is accompanied by subplots showing X and Y lineouts of the simulated data (blue) and the fit (black). 
Results of the primary simulations are shown in Fig.~\ref{fig:Si331}(b). The reflector's half-apertures $\omega_0 = 25$~mm and $l_0 = 10$~mm were chosen to satisfy defocusing aberration tolerance $\Delta \rho/\rho = 0.05$, which is the minimum required to resolve the mask pattern at periphery (i.e., two resolution elements per mask period). The choice of the meridional aperture defined the total energy bandwidth $\Delta E_X \simeq 500$~eV accepted by the crystal which was assigned to the x-ray source in the primary simulation. Source emission was modeled using a uniform distribution of energies within this bandwidth and a uniform x-ray divergence in the intervals slightly exceeding geometric acceptances of the crystal. 
The mask features are barely resolved at the periphery in the vertical and horizontal directions while the resolution is lost at the periphery along the diagonals as expected. We note that the intensity when averaged over several mask periods remains approximately uniform across the field of view. Thus, the apparent non-uniformity is primarily attributable to a local loss in the image contrast. The corresponding simulated image of the Gaussian source is nearly Gaussian in both dimensions and its FWHM ($\simeq$ 5.1~mm) shows only minimal broadening beyond the expected 5 $\mu$m.

To demonstrate improvement of the spatial resolution the half-apertures and the energy range were reduced by factor of 2 (i.e.,  $\omega_0 = 12$~mm, $l_0 = 5$~mm, and $\Delta E_X = 250$~eV), which corresponds to a set aberration tolerance of $\Delta \rho/ \rho = 0.025$. The corresponding image is shown in Fig.~\ref{fig:Si331}(c). The mask features are fully resolved, yet some blurring is noticeable at the periphery.  The supplementary image of the Gaussian source is in excellent agreement with the nominal shape and size of the source as in the previous case (Fig.~\ref{fig:Si331}(b)).

In the final test of the Si 331 configuration, ellipsoidal shape was replaced with a toroidal shape while preserving the simulation parameters of the preceding case with $\Delta \rho/\rho = 0.025$. The toroidal shape was constructed using the same principal radii of curvature as those of the ellipsoid.   
The simulated images are shown in Fig.~\ref{fig:Si331}(d). The mask features are significantly blurred to the extent that resolution of the individual mask features becomes problematic across the entire field of view. The supplementary image of the Gaussian source is strongly distorted by high order aberrations where strong non-Gaussian image tails extend primarily in the meridional direction. Note that the $x$ and $y$ ranges are extended by factors of about 4 and 3 respectively as compared to those of supplementary plots in Figs.~\ref{fig:Si331}(b,c)  
Such appearance of the aberrated image is described in detail by Kirkpatrick and Baez\cite{KB48} for grazing incidence mirrors. We note that while the overall extent of the aberrated image can be derived analytically(e.g., Ref.\cite{KB48}) a ray-tracing simulation reveals details about its shape. 
Our simulations clearly show that in this case of intermediate angles an image formed by the toroidal crystal may become unacceptable while an equivalent ellipsoidal reflector can resolve features of interest on the microscale. 

\begin{figure*}
\includegraphics[scale=0.5]{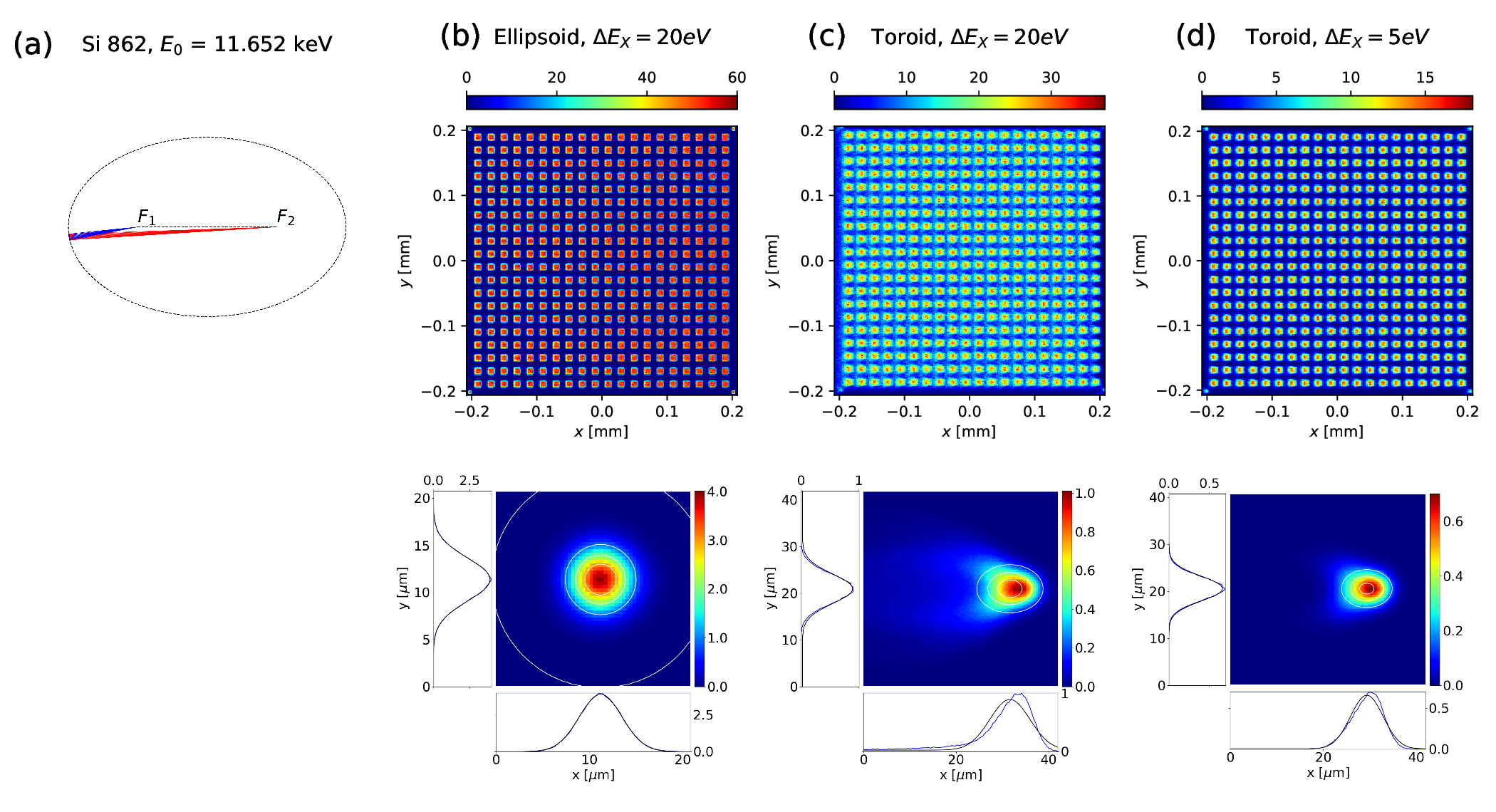}
\caption{\label{fig:Si862} 
a) Layout of simulated ellipsoidal geometry using Si 862 reflecting crystal in the meridional midplane (not to scale). 
The x-ray source plane is located at $F_1$ normal to incident rays (blue); the detector plane is placed at $F_2$ normal to reflected rays (red).
(b) Ray-traced image using ellipsoidal reflector and bandwidth of 20~eV ((parameters are shown in Table~\ref{tab:param}). 
(c) Ray-traced image using corresponding toroidal reflector and the same bandwidth of 20~eV. 
(d) Ray-traced image using the same toroidal reflector as in (c) and a reduced bandwidth of 5~eV. 
(b-c) are shown in the source/mask coordinates $x,y$. 
Images in (b-c) are shown in the source coordinates $x,y$ (i.e., coordinates normalized by the nominal magnification). 
For each case the main figure (top) corresponds to a simulation using a uniform x-ray source with a mask (Fig.~\ref{fig:mask}), and the supplemental figure (bottom) corresponds to a simulation 
using a $5~\times~5~\mu m^2$ (FWHM) Gaussian source placed in the center of the object plane. 
The sources have uniform x-ray divergences and a Gaussian distribution of photon energies with the specified bandwidths (FWHM). 
}
\end{figure*}

Simulations performed for the case of near-backscattering geometry (Si 862) schematically shown in Fig.~\ref{fig:Si862}(a) (not to scale) 
reveal additional insights.
In this case the field of view is not limited by the second order aberrations for a practically sized crystal aperture. The chosen half-aperture sizes were 30 and 20~mm in the meridional and sagittal directions, respectively. Source emission was modeled using a uniform x-ray divergence and a Gaussian distribution of energies with $\Delta E_X = 20~eV$ being the full width at half maximum (FWHM) (Table~\ref{tab:param}). We note that the extent of the crystal in the meridional direction truncates the accepted range to [$E_0-10$, $E_0+20$]~$eV$ to avoid the region near the exact backscattering, which requires more extensive simulations of x-ray reflectivity. 

The image of the masked source produced in the primary simulation using ellipsoidal crystal is shown in Fig.~\ref{fig:Si862}(b). All mask features are fully resolved while blurring of the features at the periphery as in the best case of Si 331 imager (Fig.~\ref{fig:Si331}(c)) is not observed. The supplementary simulated image of the Gaussian source placed in the center of the object plane is in excellent agreement with the nominal shape and size of the source.
The result of a simulation with the equivalent toroidal geometry using the same source bandwidth $\Delta E_X = 20~eV$~(FWHM) is shown in Fig.~\ref{fig:Si862}(c). While the features are fully resolved they appear blurred and noisy with substantial fraction of photons detected in locations nominally corresponding to completely opaque regions of the imaged mask. 
The supplementary image of the Gaussian source reveals a strong tail in the meridional direction while retaining the Gaussian shape in the sagittal direction. 
Note that the X and Y ranges are extended by factor of 2 as compared to those of the supplementary plot in Fig.~\ref{fig:Si862}(b).

In the final simulation of the near-backscattering Si 862 case the bandwidth of the source was reduced to 5~eV FWHM (Fig.~\ref{fig:Si862}(d)). The image quality has improved, yet the features are still blurred if compared to that of the primary simulation using the ellipsoidal shape (Fig.~\ref{fig:Si862}(b)). 
The supplementary image of the Gaussian source shows a considerable reduction in the meridional tail, which can be understood as aberration suppression due to reduction in the working crystal aperture due to reduction in the bandwidth of the source. 

It is illustrative to compare the last result for the near backscattering case (Fig.~\ref{fig:Si862}(d)) with simulations for a spherical crystal of the same reflection, central photon energy, working distance  and magnification. The results are presented in Appendix \ref{sec:a4}. Since the sagittal and meridional image planes are spatially separated the detection plane in the simulation was located in between the two at a distance where imaging contrast evaluated for the mask was equalized in the meridional and sagittal directions. 
The overall resolution of the mask features is considerably worse compared to the case of toroidal crystal while the extent of the Gaussian source image approximately doubles (Fig.~\ref{fig:a41}). 
This is due to the well-known astigmatic aberration in spherical reflectors present for the entire field of view.  

An additional insight emerges from the complete set of the performed simulations, where images are plotted after normalization to the energy bandwidth and the x-ray source angular divergences specified for each case. The resulting signal intensities, reported in these normalized units and indicated by the colorbars, show improved image contrast for the ellipsoidal configurations, with higher peak intensities.

\section{Preliminary assessment of the Viability of Ellipsoidal Crystal Imagers}\label{sec:viab}

\begin{figure}[!h]
\includegraphics[scale=0.5]{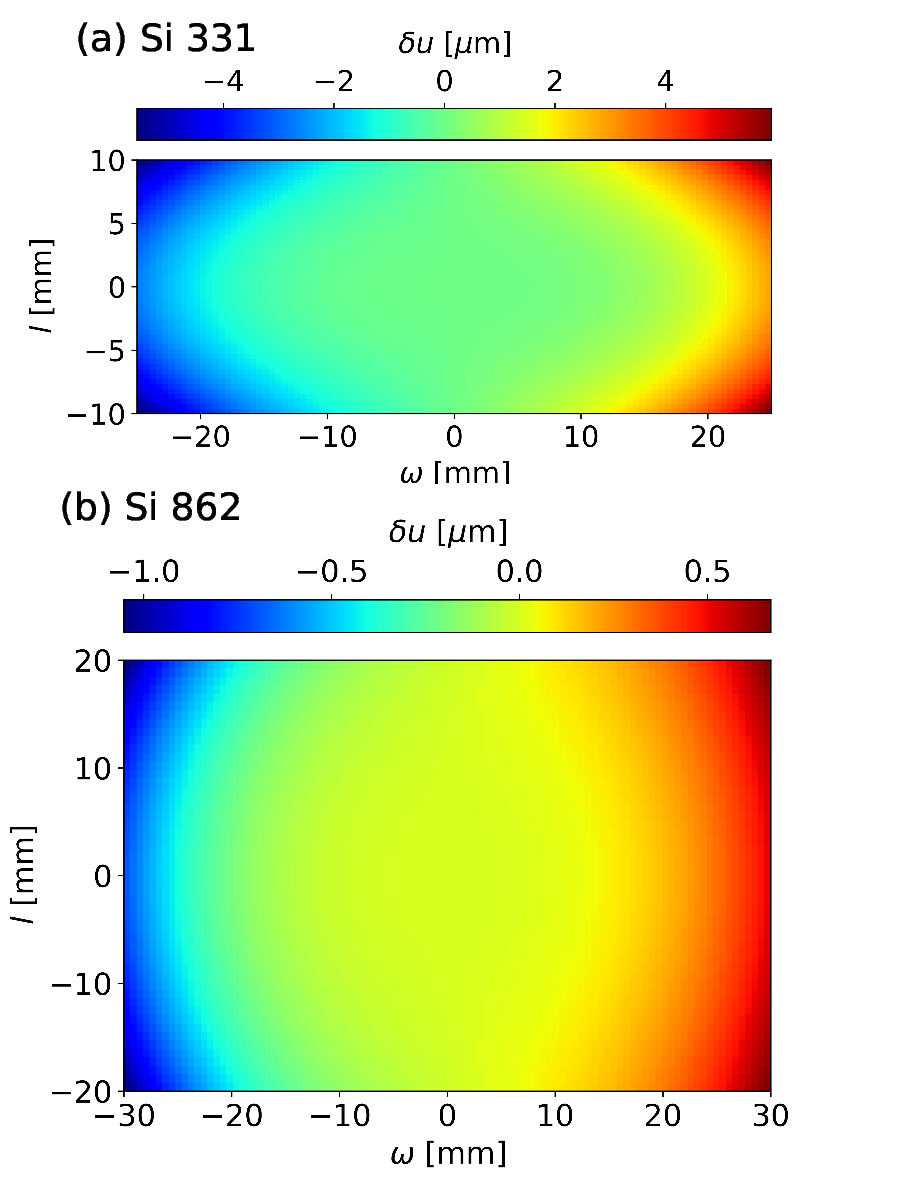}
\caption{\label{fig:resid} Maps of differences between the ellipsoidal and toroidal reflector profiles arising from higher than second order terms in the surface expansion:
(a) for the case of intermediate glancing angles (Si 331), (b) for the near backscattering case (Si 862).}
\end{figure}

Figure~\ref{fig:resid} shows maps of the differences $\delta u$ between the ellipsoidal and toroidal reflector profiles that arise from higher than second order terms in the surface expansion.
Figure~\ref{fig:resid}(a) shows the map corresponding to the studied case of intermediate glancing angles (Si 331) 
while Fig.~\ref{fig:resid}(b) shows the map corresponding to the studied near backscattering case (Si 862).  
The maximum profile differences at the reflector periphery are on the order of a few $\mu$m, requiring precision manufacturing of the aspherical substrates.
Modern high-precision surface manufacturing can achieve peak-to-valley figure errors of $\simeq$~30~nm\cite{Ebert25},
which one can adopt as a baseline estimate for reflector figure errors. 
Surface roughness on the order of 10~\AA permits direct (adhesive-free) bonding of thin parallel plate crystal slabs of thicknesses $\simeq$~50-100~$\mu m$. 
However, additional measures are needed to ensure that the crystal lattice conforms to the intended shape. 
Because the slab thickness is much larger than the required shape tolerance, the substrate surface must be engineered so that the atomic planes, 
within the effective x-ray penetration depth, follow the designed ellipsoidal profile.
Accordingly, the assurance procedure should include quantitative characterization of the curved crystal lattice, rather than surface metrology alone. 
This can be accomplished using x-ray topography in combination with x-ray diffractometry methods.\cite{Stoupin_AIPP16_2,Halavanau23}

\section{Conclusions}
In summary, an aberration-limiting form was obtained for an arbitrary-magnification imaging specular reflector with a shape approximated locally to the second order in Maclaurin surface expansion.
These defocusing aberrations arise for images of object points located away from the central ray trajectory of the system while point-to-point imaging at the center remains aberration-free due to the choice of the second order coefficients, which are defined by the meridional and sagittal radii of curvature per Coddington's equations. 
The permissible relative aperture size of the reflector (aperture acceptance size divided by the paraxial focal distance) is limited by a set relative aberration tolerance (e.g., ray aberration normalized by the size of the field of view) and scales with tangent of the Bragg angle. At high magnification, which is of interest for imaging of small objects, this condition sets a limit on the acceptance collection angle thus limiting photometric efficiency of the reflector. 
The derivation assumed no spectral bandwidth limitations and is applicable to diffracting x-ray optics with the shape of ellipsoid of revolution, 
which can provide polychromatic bandpass imaging within a range of photon energies limited by the accessible range of Bragg angles.
We find that a diffracting crystal in the symmetric Bragg geometry conformed to the ellipsoidal shape 
permits such polychromatic imaging at Bragg angles far from backscattering using aberration-limiting apertures. 
Close to backscattering the defocusing aberrations limiting the useful field of view become practically insignificant. 
Imaging properties of ellipsoidal crystals were tested using ray tracing simulations for two distinct high-magnification designs with working central Bragg angles far from backscattering 
($\theta_0~\simeq~22.5^{\circ}$) and near backscattering  ($\theta_0~\simeq~87.5^{\circ}$).  
The results of simulations are compared to those obtained with equivalent toroidal crystal geometries, defined as shapes sharing the same principal meridional and sagittal radii of curvature. In contrast to toroidal crystals, ellipsoidal crystal imagers retain high-contrast imaging performance as their acceptance aperture is increased, or equivalently as the accepted spectral bandwidth is broadened. 
This indicates that the ellipsoidal geometry suppresses higher-order image aberrations.  
Overall, our findings indicate that ellipsoidal crystal imagers can maintain performance at larger acceptance apertures, which can inform the design of next-generation high energy density science instrumentation where longer working distances lead to reduction of collection efficiency. In addition, at a fixed working distance, an increased acceptance aperture enables capture of a broader spectral band of interest, since plasma emission lines frequently have widths of $\gtrsim$10~eV. More broadly, ellipsoidal crystal imagers can benefit full-field x-ray fluorescence microscopy in applications requiring high selectivity for x-ray emission within a specified spectral band. 
 
\section{Acknowledgments}
This work was performed under the auspices of the U.S. Department of Energy by Lawrence Livermore National Laboratory under Contract DE-AC52-07NA27344.

\section{Data Availability}
The data that support the findings of this study are available from the corresponding author upon reasonable request.

\clearpage
\appendix
\section{Calculation of the Hamiltonian point characteristic}\label{sec:a1}

Upon substitution of Eqs.~(\ref{eq:surf},\ref{eq:R}) into Eqs.~(\ref{eq:main}) while omitting higher-order $u^2$ term under the square root we obtain
\begin{widetext}
\begin{subequations}\label{eq:expan}
\begin{align}
D_1 = A_1P \simeq \bigg[ p^2 \sec^2{\eta} + \omega^2 \Big(1 - \frac{p\cos{(\alpha + \epsilon)}\sin{\theta}} {2f} \Big) 
+ l^2 \Big(1 - \frac{p\cos{(\alpha + \epsilon)}}{2f\sin{\theta}}\Big) + 2 \omega p \sin{(\alpha+\epsilon)} - 2l p\tan{\eta} \bigg]^{1/2} \\
D_2 = PA_2 \simeq \bigg[ q^2 \sec^2{\eta} + \omega^2 \Big(1 - \frac{q\cos{(\alpha + \epsilon)}\sin{\theta)}} {2f} \Big) 
+ l^2 \Big(1 - \frac{q\cos{(\alpha + \epsilon)}}{2f\sin{\theta}}\Big) - 2 \omega q \sin{(\alpha+\epsilon)} + 2l q\tan{\eta} \bigg]^{1/2}
\end{align}
\end{subequations}
\end{widetext}

To expand these to the second order derivatives are calculated

\begin{subequations}\label{eq:deriv}
\begin{align}
\frac{\partial D_1}{\partial \omega} &= \frac{1}{D_1}\bigg[ \omega \Big(1 - \frac{p \cos{(\alpha + \epsilon)}\sin{\theta}}{2f} \Big) + p \sin{(\alpha + \epsilon)}\bigg] \nonumber \\ 
\frac{\partial D_1}{\partial \l} &= \frac{1}{D_1}\bigg[ l \Big(1 - \frac{p \cos{(\alpha + \epsilon)}}{2f\sin{\theta}} \Big) - p \tan{\eta}\bigg] \nonumber \\ 
\frac{\partial^2 D_1}{\partial \omega^2} &= \frac{1}{D_1} \bigg[ 1 - \frac{p \cos{(\alpha + \epsilon)}\sin{\theta}}{2f} - \Big(\frac{\partial D_1}{\partial \omega}\Big)^2 \bigg] \nonumber \\
\frac{\partial^2 D_1}{\partial \l^2} &= \frac{1}{D_1}    \bigg[ 1 - \frac{p \cos{(\alpha + \epsilon)}}{2f\sin{\theta}} - \Big(\frac{\partial D_1}{\partial l}\Big)^2  \bigg]   \nonumber \\
\frac{\partial^2 D_1}{\partial \omega \partial \l} &= -\frac{1}{D_1}\frac{\partial D_1}{\partial \omega}\frac{\partial D_1}{\partial \l}
\end{align}
\end{subequations}

Similarly,

\begin{subequations}\label{eq:deriv}
\begin{align}
\frac{\partial D_2}{\partial \omega} &= \frac{1}{D_2}\bigg[ \omega \Big(1 - \frac{q \cos{(\alpha + \epsilon)}\sin{\theta}}{2f} \Big) - q \sin{(\alpha + \epsilon)}\bigg] \nonumber \\ 
\frac{\partial D_2}{\partial \l} &= \frac{1}{D_2}\bigg[ l \Big(1 - \frac{q \cos{(\alpha + \epsilon)}}{2f\sin{\theta}} \Big) + q \tan{\eta}\bigg] \nonumber \\ 
\frac{\partial^2 D_2}{\partial \omega^2} &= \frac{1}{D_2} \bigg[ 1 - \frac{q \cos{(\alpha + \epsilon)}\sin{\theta}}{2f} - \Big(\frac{\partial D_2}{\partial \omega}\Big)^2 \bigg] \nonumber \\
\frac{\partial^2 D_2}{\partial \l^2} &= \frac{1}{D_2}    \bigg[ 1 - \frac{q \cos{(\alpha + \epsilon)}}{2f\sin{\theta}} - \Big(\frac{\partial D_2}{\partial l}\Big)^2  \bigg]   \nonumber \\
\frac{\partial^2 D_2}{\partial \omega \partial \l} &= -\frac{1}{D_2}\frac{\partial D_2}{\partial \omega}\frac{\partial D_2}{\partial \l}
\end{align}
\end{subequations}

Taylor expansions to the second order in $\omega$ and $l$:

\begin{widetext}
\begin{subequations}\label{eq:taylor}
\begin{align}
D_1 &\simeq p\sec{\eta} + \omega \frac{\sin{(\alpha + \epsilon)}}{\sec{\eta}} - l\sin{\eta} + \frac{\omega^2}{2 p \sec{\eta}}\Big[1 - \frac{p \cos{(\alpha + \epsilon)}\sin{\theta}}{2f} 
- \frac{\sin^2{(\alpha + \epsilon)}}{\sec^2{\eta}} \Big] \nonumber \\
&+ \frac{l^2}{2 p \sec{\eta}}\Big[1 - \frac{p \cos{(\alpha + \epsilon)}}{2f\sin{\theta}} - \frac{\tan^2{\eta}}{\sec^2{\eta}} \Big] 
+ \frac{\omega l}{p \sec^3{\eta}} \sin{(\alpha + \epsilon)} \tan{\eta},  \label{eq:taylor:a} \\
D_2 &\simeq q\sec{\eta} - \omega \frac{\sin{(\alpha + \epsilon)}}{\sec{\eta}} + l\sin{\eta} + \frac{\omega^2}{2 q \sec{\eta}}\Big[1 - \frac{q \cos{(\alpha + \epsilon)}\sin{\theta}}{2f} 
- \frac{\sin^2{(\alpha + \epsilon)}}{\sec^2{\eta}} \Big] \nonumber \\
&+ \frac{l^2}{2 q \sec{\eta}}\Big[1 - \frac{q \cos{(\alpha + \epsilon)}}{2f\sin{\theta}} - \frac{\tan^2{\eta}}{\sec^2{\eta}} \Big] 
+ \frac{\omega l}{q \sec^3{\eta}} \sin{(\alpha + \epsilon)} \tan{\eta} \label{eq:taylor:b} 
\end{align}
\end{subequations}
\end{widetext}

The sum of these two terms is Eqs.~\ref{eq:V123}. 

\clearpage
\section{Calculated reflectivity curves of toroidally curved Si crystals}\label{sec:a2}
\begin{figure}[!h]
\includegraphics[scale=0.4]{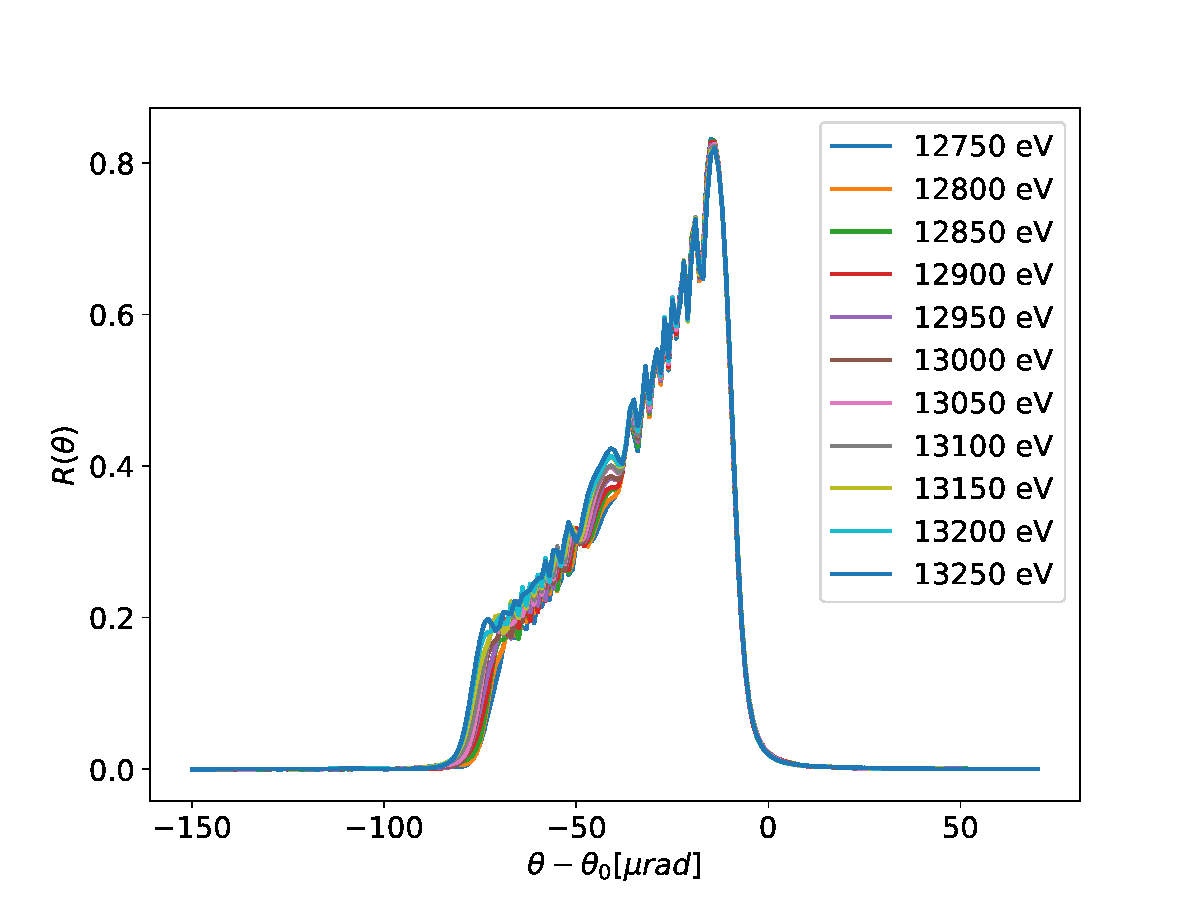}
\caption{\label{fig:a21} 
Angular reflectivity curves for Si 331 toroidally curved crystal calculated using pyTTE code.}
\end{figure}

\begin{figure}[!h]
\includegraphics[scale=0.4]{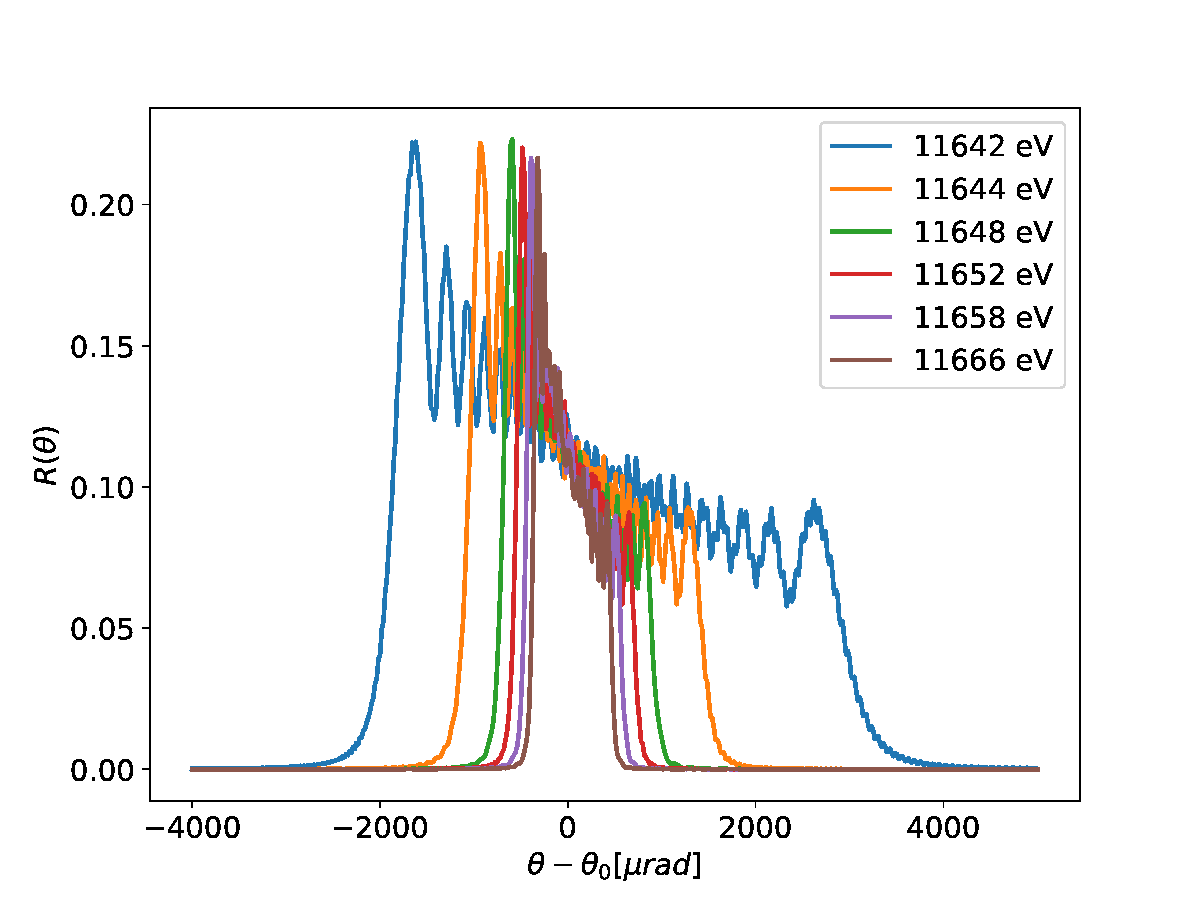}
\caption{\label{fig:a22} 
Angular reflectivity curves for Si 862 toroidally curved crystal calculated using pyTTE code.}
\end{figure}

Angular reflectivity curves were calculated using pyTTE code for Si 311 and Si 862 toroidal crystals for a grid of photon energies within the corresponding working energy intervals for each imager. 
Calculations were performed using crystal thicknesses of 100~$\mu$m in each case and the principal radii of curvature as summarized in Table ~\ref{tab:param}.
The curves averaged for $\sigma$ and $\pi$ polarizations are shown in Fig.~\ref{fig:a21} for Si~331 and in Fig.~\ref{fig:a22} for Si~862. 
The reflectivity curves are substantially broadened with respect to those of perfect flat crystals due to curvature-induced deformation of the crystal lattice. This leads to enhancement of the integrated reflectivity in the photometric formulas (Eqs.~(\ref{eq:ge},\ref{eq:g0},\ref{eq:ge_})). 
For the case of Si~331 (far from backscattering) the curves and thus the angular-integrated reflectivity $R_{\theta}$ do not vary by much within the working energy interval. 
In the other case of Si~862 (near backscattering) the variation is substantial, however is not as dramatic for the combined quantity that enters the photometric formulas $R_e = R_{\theta}dE/d\theta = R_{\theta}E/\tan{\theta}$ due to rapidly increasing $\tan{\theta}$ as the Bragg angle approaches $\pi/2$. 

\section{Maclaurin surface expansions:  ellipsoid of revolution and torus}\label{sec:a3}
The series expansion in the system of coordinates with the origin at the center of the reflector as defined in section \ref{sec:rayab}: 
\begin{equation}
u = \sum_{i = 0}^{\infty} \sum_{j=0}^{\infty} C_{ij} \omega^i l^j 
\end{equation}

As derived by Goldberg\cite{Goldberg22} for ellipsoid of revolution, limited to within the combined order $i + j = 4$ and using our notation:

\begin{subequations}\label{eq:ell_expan}
\begin{align}
C_{02} &= \frac{1}{4f \sin{\theta_0}}, \\
C_{20} &= \frac{\sin{\theta_0}}{4f}, \\
C_{12} &= \frac{(p-q)\cos{\theta_0}}{8fpq\sin{\theta_0}}, \\
C_{30} &= \frac{(p-q)\cos{\theta_0}\sin{\theta_0}}{8fpq}, \\
C_{22} &=\frac{3(p-q)^2\cos^2{\theta_0}+4pq}{32fp^2q^2\sin{\theta_0}}, \\
C_{04} &=\frac{(p-q)^2\cos^2{\theta_0}+4pq}{64fp^2q^2\sin^3{\theta_0}}, \\
C_{40} &=\frac{[5\cos^2{\theta_0}(p-q)^2+4pq]\sin{\theta_0}}{64fp^2q^2}.
\end{align}
\end{subequations}

For a torus with a minor radius $r = R_s$ and a major radius $R$ ($R_m = R + r$) the series expansion to the combined order $i+j = 6$ is\cite{bmad_manual}:
\begin{subequations}\label{eq:tor_expan}
\begin{align}
C_{20} &= \frac{1}{2R_m}, \\
C_{40} &= \frac{1}{8R_m^3}, \\
C_{60} &= \frac{1}{16R_m^5}, \\
C_{02} &= \frac{1}{2R_s}, \\
C_{04} &= \frac{1}{8R_s^3}, \\
C_{06} &= \frac{1}{16R_s^5}, \\
C_{22} &= \frac{1}{4R_sR_m^2}, \\
C_{42} &= \frac{3}{16R_sR_m^4}, \\
C_{24} &= \frac{2R_s+R_m}{16R_s^3R_m^3}.
\end{align}
\end{subequations}

If the principal radii of curvature $R_s$ and $R_m$ satisfy Eqs.~(\ref{eq:R}) the second-order terms of the two expansions are identical. 

\section{Ray tracing simulation of a spherical crystal in near backscatering geometry}\label{sec:a4}

\begin{figure}[!h]
\includegraphics[scale=0.5]{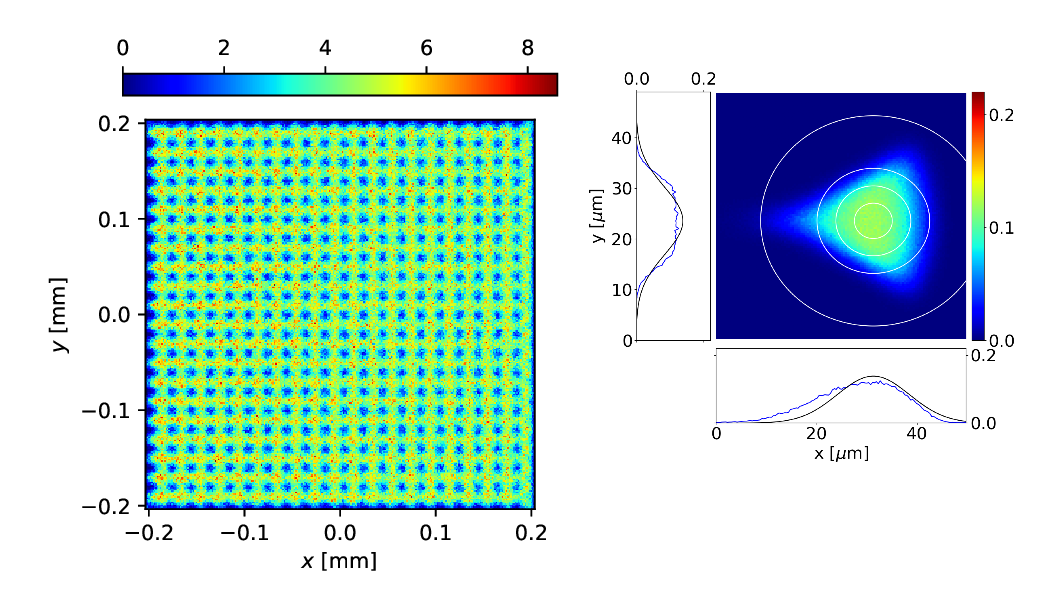}
\caption{\label{fig:a41} 
Ray-traced images using a spherical Si 862 crystal with radius of curvature $R$ = 0.929~m. The detector is placed at the distance where spatial resolutions in the 
meridional and sagittal directions are approximately equal. The simulation parameters are the same as those for the toroidal crystal with the spectral band $\Delta E_X$ = 5 eV (Fig.~\ref{fig:Si862}(d)).
Left: image of the uniform source with the mask (Fig.~\ref{fig:mask}). Right: image of the 5x5 $\mu m^2$ Gaussian source. }
\end{figure}

Figure \ref{fig:a41} shows ray-traced images for a spherical Si862 crystal with a radius of curvature equal to the meridional principal radius of curvature 
of the ellipsoidal and toroidal reflectors simulated in the Si862 near backscattering case. 
For a given working distance $p_0$, the distance between the meridional and sagittal image planes can be found as

\begin{equation}
\Delta q = \frac{q \cos^2{\theta_0}(M+1)}{\sin^2{\theta_0} - M \cos^2{\theta_0}} \simeq \Theta_0^2(M+1),
\end{equation} 
where $M = q_0/p_0$ is the magnification in the meridional plane, and $\Theta_0 = \pi/2 - \theta_0$ is assumed to be small in near backscattering.  

The above equation suggests that for a given design (fixed M and $\theta$) any two marginal rays separated by an angular distance e.g., $\Delta \beta = \omega/q$ 
will lead to an astigmatic aberration, which in the object plane can be approximated by

\begin{equation}
\Delta x \simeq \frac{\Delta \beta \Delta q}{M} \simeq \gamma \omega \Theta_0^2 \frac{M+1}{M}, 
\end{equation}
where $\gamma$ is a constant that depends on the position of the detection plane ($\gamma = 1$ corresponds to detector placed to the sagittal image plane). 
 
The detector in the simulations was placed at a distance where spatial resolution in the meridional and sagittal directions was approximately equal (based on the contrast of observed features). 
The simulation parameters were the same as those for the toroidal crystal with the spectral band $\Delta E_X$ = 5 eV (Fig.~\ref{fig:Si862}(d)).
Thus, Fig.~\ref{fig:a41} illustrates the degree of image degradation from the spherical imager’s astigmatic aberrations, reduced by placing the detector at its optimal position. 

\bibliographystyle{apsrev4-2}
\bibliography{references}

\end{document}